\documentclass[10pt]{article}
\usepackage{fullpage}
\usepackage{amsmath}
\usepackage{amssymb}
\usepackage{chemarr}
\usepackage{times}
\usepackage{epsfig,graphics}
\usepackage{graphicx}
\usepackage{subfigure}
\usepackage{epsf}
\usepackage{caption}
\usepackage{color}
\usepackage{sectsty}
\sectionfont{\large}
\newcommand*{\abstracttext}{The currently existing theory of fluorescence correlation
spectroscopy
(FCS) is based on the linear fluctuation theory originally 
developed by Einstein, Onsager, Lax, and others as 
a phenomenological approach to equilibrium fluctuations in 
bulk solutions.  For mesoscopic reaction-diffusion systems 
with nonlinear chemical reactions among a small number of 
molecules, a situation often encountered in single-cell
biochemistry, it is expected that FCS time correlation functions
of a reaction-diffusion system can deviate from the classic results 
of Elson and Magde [{\em Biopolymers} (1974) 13:1-27].   
We first discuss this nonlinear effect for reaction systems 
without diffusion.
For nonlinear stochastic reaction-diffusion systems there are
no closed solutions; therefore, stochastic Monte-Carlo simulations
are carried out. We show that the deviation is small for
a simple  bimolecular reaction; the most significant
deviations occur when the number of molecules is small and of the same order.
Our results show that current linear FCS theory could be adequate for 
measurements on biological systems that contain many other sources of 
uncertainties. At the same time it provides a framework for future 
measurements of nonlinear, fluctuating chemical reactions with high-precision 
FCS. Extending Delbr\"{u}ck-Gillespie's theory for stochastic 
nonlinear reactions with rapidly stirring to 
reaction-diffusion systems provides a mesoscopic model 
for chemical and biochemical reactions at nanometric and 
mesoscopic level such as a single biological cell.}

\let\oldmaketitle\maketitle
\let\maketitle\relax

\title{Fluorescence Correlation Spectroscopy and
Nonlinear Stochastic Reaction-Diffusion}

\author{Mauricio J. Del Razo\textdagger, Wenxiao Pan\textdaggerdbl, Hong Qian\textdagger, and Guang Lin\textdaggerdbl \\
{\small \textit{University of Washington, Seattle, WA 98195-3925, and Pacific Northwest National Laboratory, Richland, WA 99352.}}\\[2mm]
 {\small email: maojrs@uw.edu}}

\date{}

\begin{document}
\def\vf {{\bf f}}
\def\vj {{\bf j}}
\def\vn {{\bf n}}
\def\vP{{\bf p}}
\def\vp{{\bf p}}
\def\vx{{\bf x}}
\def\vu{{\bf u}}
\def\vv{{\bf v}}
\def\vw {{\bf w}}
\def\mA{{\bf A}}
\def\mB{{\bf B}}
\def\mC{{\bf C}}
\def\mD{{\bf D}}
\def\mG{{\bf G}}
\def\mI{{\bf I}}
\def\mS{{\bf S}}
\def\mQ{{\bf Q}}
\def\mR{{\bf R}}
\def\mT{{\bf T}}
\def\mU{{\bf U}}
\def\mV{{\bf V}}
\def\w{\omega}
\def\u{\nu}
\def\mGa{\mbox{\boldmath$\Gamma$}}
\def\mPhi{\mbox{\boldmath$\Phi$}}
\def\mPi{\mbox{\boldmath$\Pi$}}
\def\mXi{\mbox{\boldmath$\Xi$}}
\def\vg {\mbox{\boldmath$\gamma$}}
\def\vpi{\mbox{\boldmath$\pi$}}
\def\vphi{\mbox{\boldmath$\phi$}}
\def\vnu{\mbox{\boldmath$\nu$}}
\def\vkappa{\mbox{\boldmath$\kappa$}}
\def\vmu{\mbox{\boldmath$\mu$}} 

\twocolumn[
\begin{@twocolumnfalse}
\oldmaketitle
\begin{abstract}
\abstracttext
\end{abstract}
\end{@twocolumnfalse}
]

\section{{Introduction}}
\noindent
Single-molecule studies of biological macromolecules
focus on conformational states of individual molecules
and transitions between 
states \cite{xie1998single,bustamante2008singulo,graslund2010single,qian2014statistics}.  
Concentration fluctuation spectroscopy, on the other hand, 
measures the molecular number fluctuations associated with 
linear, and nonlinear, biochemical reactions \cite{elson1975concentration,weissman1981fluctuation}.  
For unimolecular reactions,
these two approaches are conceptually equivalent, in 
statistical terms, via the
multi-nomial distribution: If a single molecule has 
$K$ states with $p_k(t)$ being the probability for the
molecule in state $k$ at time $t$, then for $M$ independent
copies of the same molecule, one has the probability
distribution for $m_{\ell}$ number of molecules in $\ell$ 
state following \cite{hill2004studies,hill2003approach}
\begin{equation}
    \frac{M!}{m_1!m_2!\cdots m_K!}\left(p_1(t)\right)^{m_1}
         \left(p_2(t)\right)^{m_2}\cdots 
         \left(p_K(t)\right)^{m_K}, \ \ \ 
\end{equation} 
where $m_1+m_2+\cdots+m_K=M$.  In fact more specifically, 
if the first-order rate constant for transition $k\rightarrow
j$ is $q_{kj}$, then in the concentration fluctuation
measurements of a system with $M$ independent copies of the
molecule, the temporal correlation function is simply M times the 
correlation function derived from a single-molecule measurement.
If we denote the ``state of the reaction system'' by 
$\{m_1,m_2,\cdots,m_K\}$, the rate constant for transition from 
$\{m_1,\cdots,m_j,\cdots,m_{\ell},\cdots,m_K\}$ to  
$\{m_1,\cdots,m_j-1,\cdots,m_{\ell}+1,\cdots,m_K\}$
is $m_jq_{j\ell}$.   Experimentally, single-molecule
measurements on state fluctuations have a much more
superior signal-to-noise characteristics than the 
concentration fluctuation.  

  However, for reaction systems with nonlinear reactions 
such as $A+B\rightleftharpoons
C$, the two approaches no longer provide equivalent 
information; they are in fact complementary.
This distinction has not been widely appreciated.
Corresponding to chemistry reaction theories, the 
single-molecule approach parallels nicely with Kramers' 
reaction rate theory \cite{hanggi1990reaction}, while the concentration
fluctuation measurements is intimately related to 
Delbr\"{u}ck's chemical master equation, or 
Gillespie's stochastic kinetics, for chemical reaction 
{\em systems} with reaction networks \cite{delbruck2004statistical,gillespie2007stochastic,qian2010cellular}.
In the latter systems, rate constants for individual reactions
are supposely known {\em a priori}; complex chemical
or biochemical behavior arises as a consequence of a
nonlinear reaction network \cite{qian2014statistics}.

  Fluorescence correlation spectroscopy (FCS) is one of
the leading physiochemical techniques to experimentally 
measure concentration fluctuations of  
nonlinear chemical reactions with stochastic fluctuations 
in mesoscopic systems \cite{elson_book}.  Other methods
include conductance fluctuations for electrochemical
reaction \cite{feher1973fluctuation}.  With the newfound 
perspective given above, especially with nonlinear 
chemical reactions in
mind, we re-visit the original theory of FCS
developed by Elson and Magde (EM) \cite{elson1974fluorescence}.
We show that the EM theory is based on the 
universally valid phenomenological linear approximation approach
to macroscopic fluctuations, developed
by Einstein for Brownian motion, Onsager and Machlup
for linear Gaussian fluctuations \cite{onsager1953fluctuations}, 
and Lax for nonequilibrium steady state \cite{lax1960fluctuations}.
A systematic exposition is given by Keizer\cite{keizer1987statistical}.
The original EM theory was motivated by Eigen's linear
relaxation kinetics \cite{friesstechnique} and Onsager's 
regression hypothesis \cite{onsager1931reciprocal}.  It has 
been experimentally verified for concentration
fluctuations in bulk 
solutions \cite{fcs_prl,ehrenberg1974rotational}.

There is a growing interest in the concentration 
(or copy-number) fluctuation studies on single
live cells, both experimental  \cite{xie2008single} 
and theoretical \cite{qian2010cellular}.  In this Letter, 
we show that for some systems the linear, 
phenomenological fluctuation theory breaks down, and 
a mechanistic nonlinear stochastic reaction theory is 
necessary.  \\
	
\noindent
{\bf\em Nonlinear chemical reaction.}
The nonlinear effect we discuss for FCS is also
present in chemical relaxation kinetics.  To illustrate
this, consider the bimolecular reaction
$A+B \overset{k_f}{\underset{k_{b}}{\rightleftharpoons}} C$.
According to Eigen's theory, linear relaxation kinetics 
gives a single time constant $\tau_r$:
\begin{equation}
         \tau_r^{-1} = k_f\Big(c_A^{eq}+c_B^{eq}\Big) 
               + k_b = k_f\Big(c_A^{tot}+c_B^{tot}-2c_C^{eq}\Big) 
               + k_b.
\end{equation}
The relaxation kinetics then is a single exponential for
the concentration of $C$:
$\big(c_C(t)-c_C^{eq}\big)
=\big(c_C(0)-c_C^{eq}\big)e^{-t/\tau_r}$.  However, the 
nonlinear kinetics based on the Law of Mass Action is 
\begin{equation}
      c_C(t)-c_C^{eq} = \left\{\frac{1}
      {1-k_f\tau_r\delta_0
          \big(1-e^{-t/\tau_r}\big)}\right\} \\
            \delta_0
                 e^{-t/\tau_r},
\label{nonlaplusbeqc}
\end{equation}
with
\begin{align*}
 \delta_0 = c_C(0)-c_C^{eq} .
\end{align*}

\begin{figure}
\centering
\includegraphics[width=0.35\paperwidth]{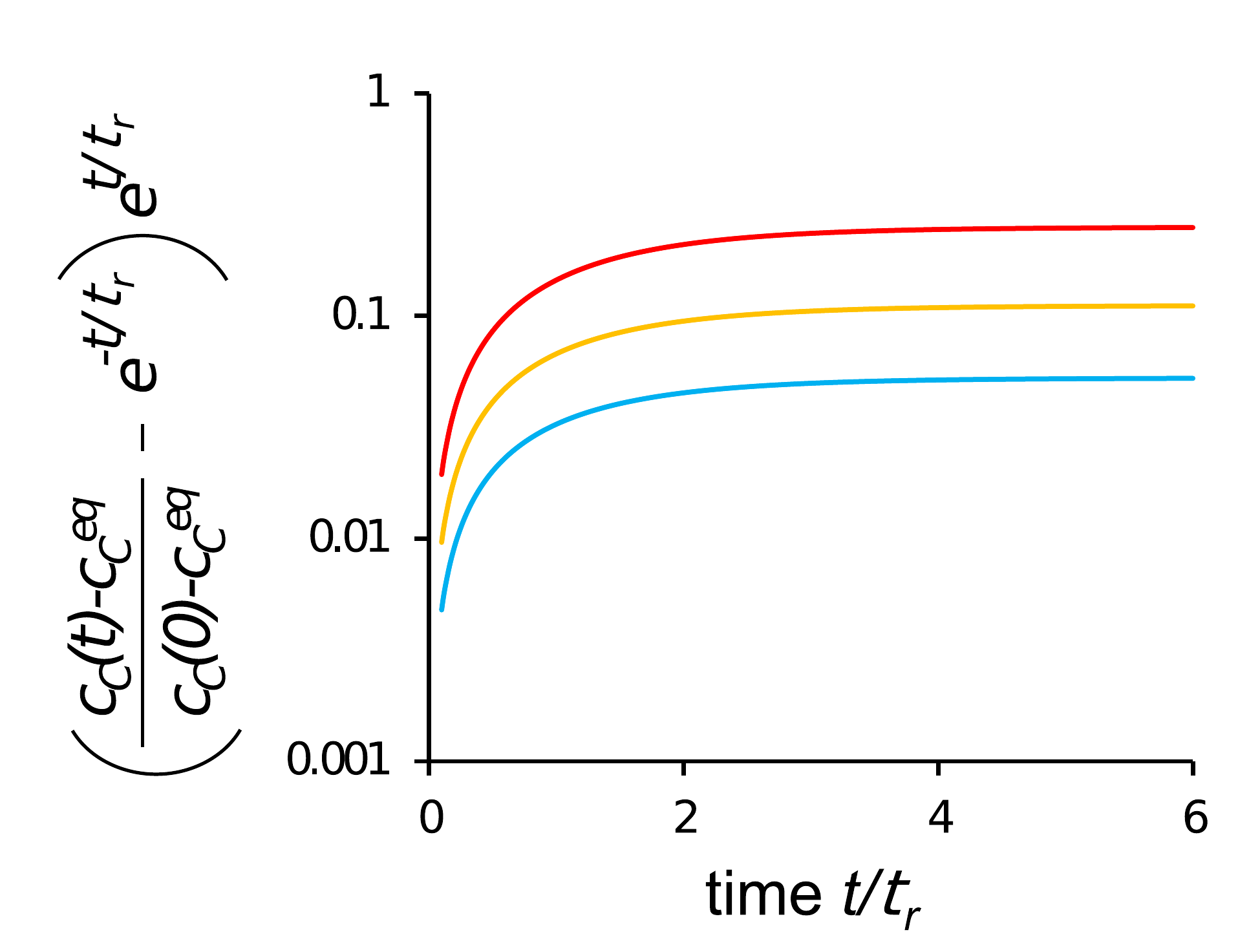}
\caption{The fractional difference between linear and 
fully nonlinear kinetics for various 
$k_f\tau_r\big(c_C(0)-c_C^{eq}\big)$ = 0.2 (red), 
0.1 (orange) and 0.05 (blue).
}
\label{fig:1}
\end{figure}

The term in the $\{\cdots\}$ is due to the nonlinear effect.
When the amplitude $\delta_0$ is sufficiently 
small, this term is negligible.  Fig. \ref{fig:1} shows 
the fractional difference between the fully nonlinear 
relaxation kinetics given in (\ref{nonlaplusbeqc}) and 
the single exponential from the linear case. 
If the amplitude $\delta_0$ is small, the fractional difference will
also be small and the linear approximation is valid; otherwise, there will 
be  a significant difference between the linear and nonlinear case.

It is important to note that for nonlinear reactions the 
relaxation time not only depends on the individual rate constants, but 
also on the concentrations at equilibrium. Furthermore, for 
nonlinear reactions, the reaction rates depend on diffusion, as it's
clearly established in the theory of diffusion
limited reactions \cite{berg1978diffusion,collins1949diffusion,szabo1980first}. Although not apparent
in the Law of Mass Action, the diffusion is necessary to 
determine the macroscopic reaction rate. The coupling between the diffusion and the reaction rates is the actual source of the nonlinearity, and it is also the reason why the relaxation time depends on the composition at equilibrium. With this realization in mind, it becomes clear that the bimolecular association rate constant $k_f$, and the diffusion constants for $A$ and $B$ used in EM theory are not independent parameters, see Appendix \ref{sec:Appendix}.  It is worth to mention that in the MesoRD approach to stochastic reaction-diffusion, these parameters are treated as independent \cite{hattne2005stochastic}.

For stochastic reaction-diffusion 
kinetics, the difference between the linear and nonlinear systems
can also be significant; however,  the FCS systems are sufficiently complex 
and no analytical results are available. In order to address this issue, the
present study will rely on Monte Carlo simulations to study nonlinear
reaction-diffusion systems.

In Sec. \ref{sec:methods}, the 
implementation of an stochastic Monte Carlo method
is explained.   The parameters employed in the simulations 
are shown to be consistent with the ones used in EM theory. 
Sec. \ref{secc:results} shows the 
comparisons between the mesoscopic, nonlinear
correlations obtained from Monte Carlo simulations and the 
EM theory. The discrepancies 
between these two are further discussed in section \ref{sec:discuss}.
For completeness, an appendix with the analytical results of the EM theory is 
included.

\section{Simulation Methods}
% ADD CONNECTION TO GILLESPIE ALGORITHM\\
\label{sec:methods}
\noindent
The results from the simulations performed in the present work will be compared to the results 
from EM theory \cite{elson1974fluorescence}.  Since some of the 
key molecular parameters used in 
the simulations are different from the ``macroscopic'' rate constants and diffusion
coefficients in the EM theory, it is important that the 
parameters employed in the simulation are consistent with those from EM theory.
We explain how the simulation was performed and show that the parameters are consistent.

Three-dimensional diffusion is simulated in terms of 
3D random walk with time step, $\Delta t$, and length step, $\epsilon_\xi$.
These parameters are related to the 3D diffusion constants used in EM theory $D_\xi=\epsilon_\xi^2/\big(6\Delta t\big)$ with $\xi=A,B$ or $C$. The reason for different step sizes is to keep all the particles moving with a single frequency, which means, same time step and different diffusion coefficients. Initially, particles are positioned randomly with uniform distribution in a cubic box of dimensionless length 2$\times$2$\times$2 with periodic boundary conditions. There are three molecular species 
represented by particles, $A$, $B$, and $C$. At every time step, each particle moves with a distance of $\epsilon_\xi$ in one of
the $\pm x$, $\pm y$ and $\pm z$ directions at equal probability of 
$1/6$. Depending on if a reaction is involved, we have three cases to be discussed below in detail.  \\

\noindent 
{\bf\em Pure diffusion.}
In this case, since reactions are not involved, there is only one specie (say $A$) of particles moving in a 3D random walk. In order to emulate FCS in simulation, we introduce a laser beam in the transverse plane with a cylindrical Gaussian intensity profile\cite{elson1975concentration},
\begin{equation}
I(\mathbf{r}) = I_0 e^{-2(x^2+y^2)/w^2},
\label{equ:laser_intensity}
\end{equation}
where $I(\mathbf{r})$ is the intensity of the incident laser light at position $\mathbf{r}=(x,y)$, $I_0$ is the maximum intensity at the center of the beam and the focal volume $w$ is given by the radius at which $I/I_0=e^{-2}\ll 1$.  In the pure diffusion, as well as in all the other cases, the focal volume $w$ needs to be much smaller than the simulation box and considerably larger than the length step. In this case, setting the value of $w$ to the length step ($\epsilon_A$) as $w=10\epsilon_A$ is sufficient for good accuracy. The fluctuation in the photocurrent is 
caused by the concentration fluctuation through 
\begin{equation}
\delta I(t) = I(t) - \langle I(t)\rangle,
\label{equ:intensity_fluc}
\end{equation}

with,
\begin{equation*}
  I(t) = I_0 \sum_{i=1}^{N_A} e^{-2(x_i^2+y_i^2)/w^2}.
\end{equation*}

Here, $N_A$ is the total number of $A$ particles and $x_i,y_i$ their positions. Therefore, the correlation of concentration fluctuations is just the temporal autocorrelation of the photocurrent, $G(\tau)$, calculated in the simulation as
\begin{equation}
G(\tau) = \frac{1}{N_t-j} \sum_{k=1}^{N_t-j}\delta I\big(k\Delta t\big)\delta I\big((k+j)\Delta t\big),
\label{equ:autocorr_fluc}
\end{equation}
where $\tau=j\Delta t$, $j$ starts from 0, and $N_t$ is the total number of time steps. $G(\tau)$ is further normalized dividing by $G(0)$.
Note we are treating the molecules as point-like light sources; however, the molecular radii will be relevant when calculating the binding radius for the bimolecular reaction case.\\

\noindent
{\bf\em Unimolecular isomerization.}
In this case, we consider a reaction of type
$A \xrightleftharpoons[k_b]{k_f} B$
with rate constants $k_f$ and $k_b$, and an equilibrium constant $K_{eq}=k_f/k_b$. In the simulation, we set $k_f$ and $k_b$ as rate parameters of exponentially distributed waiting times. These are related to the probability of the reactions occurring via 
\begin{align*}
P_{A\rightarrow B}=1-\exp{(-k_f\Delta t)}, \\
P_{B\rightarrow A}= 1-\exp{(-k_b \Delta t)},
\end{align*}
respectively \cite{gillespie2007stochastic}. Without loss of generality, we assume particle $A$ and $B$ have the same diffusion coefficient and thereby the same characteristic length step $\epsilon$.  Once again, it's sufficient to choose the focal volume $\omega=10\epsilon$. At each time step, besides of performing a random walk, particle $A$ can become $B$ with probability $P_{A\rightarrow B}$, and particle $B$ can become $A$ with probability $P_{B\rightarrow A}$. Concentration fluctuations for $A$ or $B$ are measured analogously to the pure diffusion case. \\

\noindent
{\bf\em Bimolecular reaction.}
Some very effective and accurate methods have been developed for the bimolecular reaction with diffusion when the number of ligands is large \cite{popov2001three, popov2001three2}. These are based on the analytical solution for the reversible diffusion-influenced reaction for an isolated pair in 1D and 3D \cite{kim1999exact,agmon1984diffusion,gopich1999excited,kim1999excited}. Unfortunately, the current work will require simulations in the nonlinear regime where we have a small number of ligands where these methods might not be appropriate. There are also other popular tools like the MesoRD and Smoldyn software to simulate these and other type of reactions \cite{hattne2005stochastic,andrews2004stochastic}; however, the simplicity of the simulations required allowed us to produce our own code employing a similar approach to Smoldyn\cite{andrews2004stochastic}.

Before addressing how to perfom the simulations for bimolecular reactions, note EM theory calculates the final autocorrelation function for the bimoleculer case as the sum of all the correlations weighted accordingly,
\begin{align*}
G(\tau) = \sum_{j=1}^{m}\sum_{l \le j} (2 - \delta_{jl})G_{jl}(\tau),
\end{align*}
where $G_{jl}$ is the correlation between molecule $j$ and $l$, and $j,l$ can be $A,B$ or $C$. The function $G(\tau)$ contains the resulting photocurrent correlations from all the fluorescent molecules $A$, $B$ and $C$, including coupling effects. The autocorrelation function $G(\tau)$ is the one that is actually compared to real experiments because it's not experimentally possible to isolate the fluorescence of $A$ from that of the reaction product $C$. However, in our computational setting, we can allow ourselves to concentrate on only one of these correlation curves, $G_{CC}(\tau)$. This curve obtained from only the photocurrent fluctuations of molecule $C$ contains all the information we need, including both reaction rates. As EM theory should remain consistent, the error made when calculating the reaction rates by fitting the simulation correlation curve with the theoretical $G_{CC}(\tau)$ are equivalent to those made when fitting the experimental curve with the theoretical $G(\tau)$, with the exception of additional experimental errors. How to obtain the autocorrelation function $G_{CC}(\tau)$ that we will compare to our simulations is shown in the Appendix \ref{sec:Appendix}.

For the bimolecular simulations, the reaction is assumed to be 
$A+B\xrightleftharpoons[k_b]{k_f} C$
with second-order rate constant $k_f$ (forward reaction rate), first-order 
rate constant $k_b$ (backward reaction rate) and equilibrium constant 
$K_{eq}=k_f/k_b=c_c^{eq}/\big(c_A^{eq} c_B^{eq}\big)$. Here, $c^{eq}_{\xi}$ is the equilibrium concentration of 
specie $\xi$.  We assumed the diffusion 
coefficients to be $D_A=D_C=D$ and $D_B \ge D$, since usually $A$ and $C$ are considered macromolecules and $B$ plays the role of a small ligand.  Consequently, the lengths steps will obey $\epsilon_A=\epsilon_C \le \epsilon_B$, and a sufficiently accurate focal volume is found to be $\omega=25\epsilon_B$. Note the diffusion constant
of a particle is not determined by its molecular weight {\em per se}
but by its hydrodynamic radius. We set the probability of a backward reaction to occur in terms of the parameter $\kappa_b$ as,
\begin{align}
P_{C\rightarrow AB} = 1- \exp{(-\kappa_b\Delta t)}.
\label{equ:probabilities}
\end{align}

At every ${\Delta t}$ we check if any reaction occurs. For the forward reaction, we assume it's diffusion limited. When the distance between molecules $A$ and $B$ is less than the binding radius $R$, they react with probability one. The binding radius is given by the sum of the radii of $A$ and $B$. 
 Note that the binding radius should be much smaller than the focal volume $\omega$, this condition is imposed in all the simulations.
In the backward reaction, particle $C$ becomes $A$ and $B$ simply with probability $P_{C\rightarrow AB}$. The newly formed $A$ molecule is placed where the $C$ molecule was, and the $B$ particle is placed a distance $R_u$ away from it in a random direction, with $R_u>R$.  If the $B$ particle happens to be placed inside the binding radius of another $A$ molecule, another random direction is chosen to avoid an artificial binding. The introduction of an unbinding radius $R_u$ is a possible solution to simulate the many-particle reaction accurately and address the issue of geminate recombinations in the diffusion limited model \cite{andrews2004stochastic}.

Note we called the backward rate $\kappa_b$ and not $k_b$. As particles $A$ and $B$ need to collide first before reacting, the effective forward rate $k_f$ required to compare to EM theory is unknown. Consequently, it is not clear if $\kappa_b$ should be the effective backward rate $k_b$ either. Also note that for the bimolecular case, we do not expect the macroscopic concentrations relaxation to be exponential but a power law \cite{gopich2002kinetics}, this confirms that $\kappa_b$ from equation (\ref{equ:probabilities}) might not correspond to the effective backward rate $k_b$. This is a subtle matter that will be treated in an upcoming manuscript. For the purpose of the current work, understanding some of the dynamics of geminate recombinations\cite{agmon1990theory,khokhlova2012comparison} will help us address this issue. \\

%  lets recall the analytic results of Smoluchowski and Collins and Kimball \cite{collins1949diffusion} for the forward reaction $A+B\xrightarrow[]{k_f} C$. Both models employ a single $A$ molecule fixed centered at the origin and solve for the concentration gradient of $B$ in three dimensions using an absorbing or partially absorbing boundary condition respectively \cite{shoup1982role}. Using our simulation without a backward reaction, we reproduced these models and calculated the effective forward rate $k_f$ in terms of the first passage time $\tau$. The simulation was validated, since the results were accurate with less than $3\%$ percentage error with respect to the analytic value. However, if the backward reaction with rate $\kappa_b$ is included, the simulation no longer reproduces any of the analytic results for the forward rate. It becomes evident that the presence of backward reactions affects the forward rate. In the next paragraphs, we will show geminate recombinations are the ones responsible for this effect, and how 
% we can calculate the right forward and backward rate. \\

\noindent
{\bf\em Geminate recombinations.}
Geminate recombinations occur when a particle $B$ that just dissociated from a certain $A$, associates again with it\cite{khokhlova2012comparison,agmon1990theory,andrews2004stochastic}. At first sight, it is not evident how this phenomena alters the reaction rates. One way to understand it is in terms of the waiting times. For the first reaction, the $B$ molecule is positioned randomly with a uniform distribution outside of the reaction sphere of radius $R$. The mean first passage waiting time for the reaction to occur is the one given by Collins and Kimball's or Smoluchowski's theory\cite{collins1949diffusion,shoup1982role}. However, whenever a dissociation occurs, the $B$ particle is always positioned very near $A$.   As a result, the distribution of the initial position of the $B$ molecule should not be uniform, and the average waiting time for the forward reaction to occur again will no longer be the one given by Collins and Kimball or Smoluchowski's theory. As the average waiting times and 
the rates are inversely proportional, the effective forward rate cannot be expected to be the same either \cite{szabo1980first,shoup1982role}. In other words, the just dissociated $B$ has a higher chance to bind to the same $A$.  This is conceptually prevented in the classical Law of Mass Action, and Delbruck-Gillespie theory, which require a rapid stirring reaction vessel \cite{gillespie2007stochastic}.

So how to map the correct forward rate to EM theory? Let us focus on the Smoluchowski approach and call $\phi$ the probability of a geminate recombination to occur. In a system in equilibrium with several molecules, the law of large numbers tells us that a fraction $\phi$ of all the forward reactions are due to geminate recombinations. Therefore, we expect that the remaining fraction $1-\phi$ follows the forward reaction rate for irreversible reactions given by Smoluchowski's theory. This yields the relation between the irreversible rate of Smoluchowski\cite{collins1949diffusion}, $k_D = 4\pi R D_*$ with $D_* = D+D_B$, and the reversible effective forward rate $k_f$ as\cite{andrews2004stochastic}
\begin{align}
  k_f = \frac{k_D}{1-\phi}.
  \label{revtoirr}
\end{align}
In order to calculate the backward rate, we only need to know the equilibrium constant $K_{eq}$. From our simulation, we can calculate it in terms of the average concentrations as $K_{eq}=c_c^{eq}/\big(c_A^{eq} c_B^{eq}\big)$. However, we also know $K_{eq}= k_f/k_b$, so the effective backward rate can be calculated as
\begin{align}
  k_b = k_f/K_{eq}.
  \label{backrate}
\end{align}
The only question left to answer is how to calculate the probability of geminate recombinations $\phi$. As geminate recombinations are actually an stochastic process, there are many issues to deal with in order to fully address that question. A separated manuscript regarding some of these is in preparation. However, a simple approach using Smoluchowski's original solution with an unbinding radius will be sufficient for our current purpose\cite{andrews2004stochastic}. We can solve Smoluchowski's steady state equation for a reversible reaction in equilibrium using an absorbing boundary condition at $R$ and a constant $B$ concentration at $R_u$ \cite{andrews2004stochastic}. These boundary conditions mean that the flux at the source $R_u$ equals the flux at the sink in $R$, as it's expected from a reversible reaction at equilibrium. The solution to this boundary problem is given by
\begin{align*}
  \rho(r) = c_0 \frac{R_u(R - r)}{r(R - R_u)},
\end{align*}
where $\rho(r)$ can be understood as the concentration of $B$ or as the radial distribution function if it's normalized.
The reaction rate is given by the flux,$4\pi R^2 D_*\rho'(R)$ per average concentration of $B$ molecules, $c_0$, 
\begin{align}
  k_f = \frac{4\pi R D_*}{1 - \frac{R}{R_u}} = \frac{k_D}{1 - \frac{R}{R_u}}.
  \label{frate}
\end{align}
Note that as $R_u\rightarrow \infty$ the original Smoluchowski's irreversible rate is recovered. Comparing this result with equation (\ref{revtoirr}) yields that $\phi = R/R_u$ \cite{andrews2004stochastic}. It's important
to note this analytic result is only fully valid in the limit of infinitely accurate Brownian Motion. In our simulation, we'll employ random walks with very small characteristic length step, so equation (\ref{frate}) is a good approximation of the forward rate. If the characteristic step is increased, the simulations become faster; however, a more complicated approach is necessary to calculate the correct reaction rates, as the one used by Smoldyn software\cite{andrews2004stochastic}. For the purpose of this work, computational efficiency is not an immediate issue, so we employ characteristic length steps that are small enough to use the forward rate given in equation (\ref{frate}) accurately.   

It is still not evident how to choose the appropriate unbinding radius $R_u$. An initial guess satisfying $R_u > R$ is made. Since every different unbinding radius yields different values of the forward rate, the final concentrations at equilibrium and the equilibrium constant $K_{eq}$, the simulation is executed to yield these for the initial guess. Afterwards, the unbinding radius is recalculated in terms of the final concentrations, and the process is repeated until the unbinding radius remains practically constant between iterations. The unbinding radius in each iteration is recalculated as,
\begin{align}
%   R_u = \frac{1}{2} \left[\frac{V_{box}}{nA}\right]^{1/3},
    R_u  = \left[\frac{3}{4\pi}\left(\frac{V_{box}}{nA}\right)\right]^{1/3}
  \label{UR}
\end{align}
where $V_{box}$ is the volume of the cubic box and $nA$ is the number of $A$ particles at equilibrium. This expression is obtained by assuming each $A$ molecule has it's own mini-sphere with volume $V_{box}/nA$. In this sphere we can assume the behavior is the same as in Smoluchoski's theory which involves only one copy of the $A$ molecule. Assuming $A$ is positioned at the center of the sphere , the appropriate unbinding radius would correspond to the distance between $A$ and the border of the sphere. This value is clearly given by expression (\ref{UR}).

Once the unbinding radius is calculated, $\phi$ is determined as well as the forward rate $k_f$. The effective backward rate $k_b$ is obtained with $k_f$ and the equilibrium
constant $K_{eq}$ using equation (\ref{backrate}). These last three parameters are precisely the ones used in EM theory \cite{elson1974fluorescence}. In order to validate the simulation, we tested the case with one $A$ molecule fixed at the center. The forward rate was calculated using the average first passage time after each forward reaction \cite{szabo1980first,shoup1982role} $k_f=1/(c_0\tau)$ from the simulation and the Smoluchowski equation for reversible reaction (\ref{frate}). The relative error between the two calculated forward rates was less than five percent for 100 runs with $5\times10^7$ time iterations. These results provide consistency between the parameters used in our simulation and the parameters from EM theory.

\section{Results}
\label{secc:results}
\begin{figure}
\centering
  \subfigure[]{
  	\includegraphics[width=0.45\paperwidth]{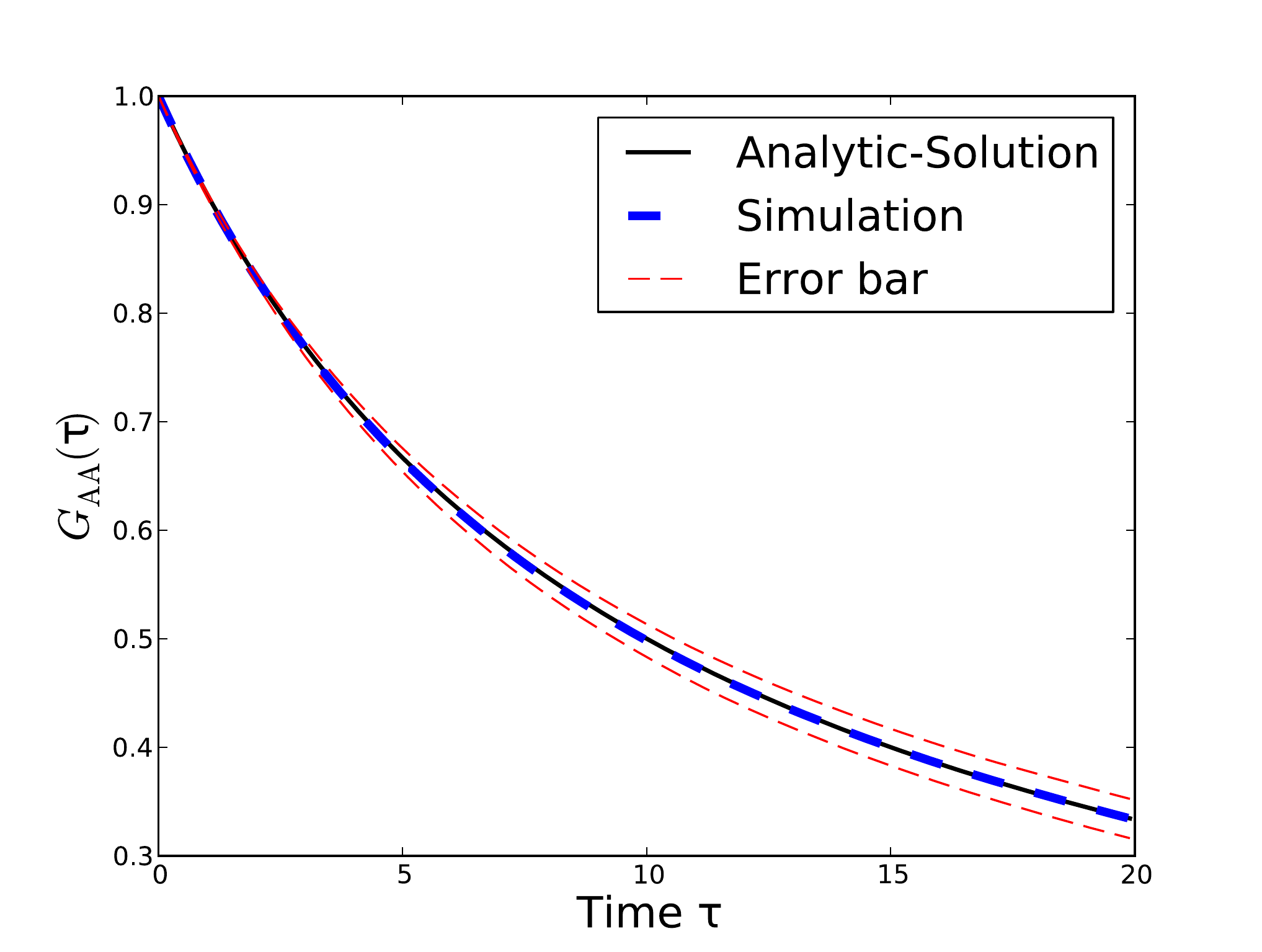} %GAA_sum.pdf}
  	\label{fig:GAA_purediff_sum-a} }
  \subfigure[]{
  	\includegraphics[width=0.45\paperwidth]{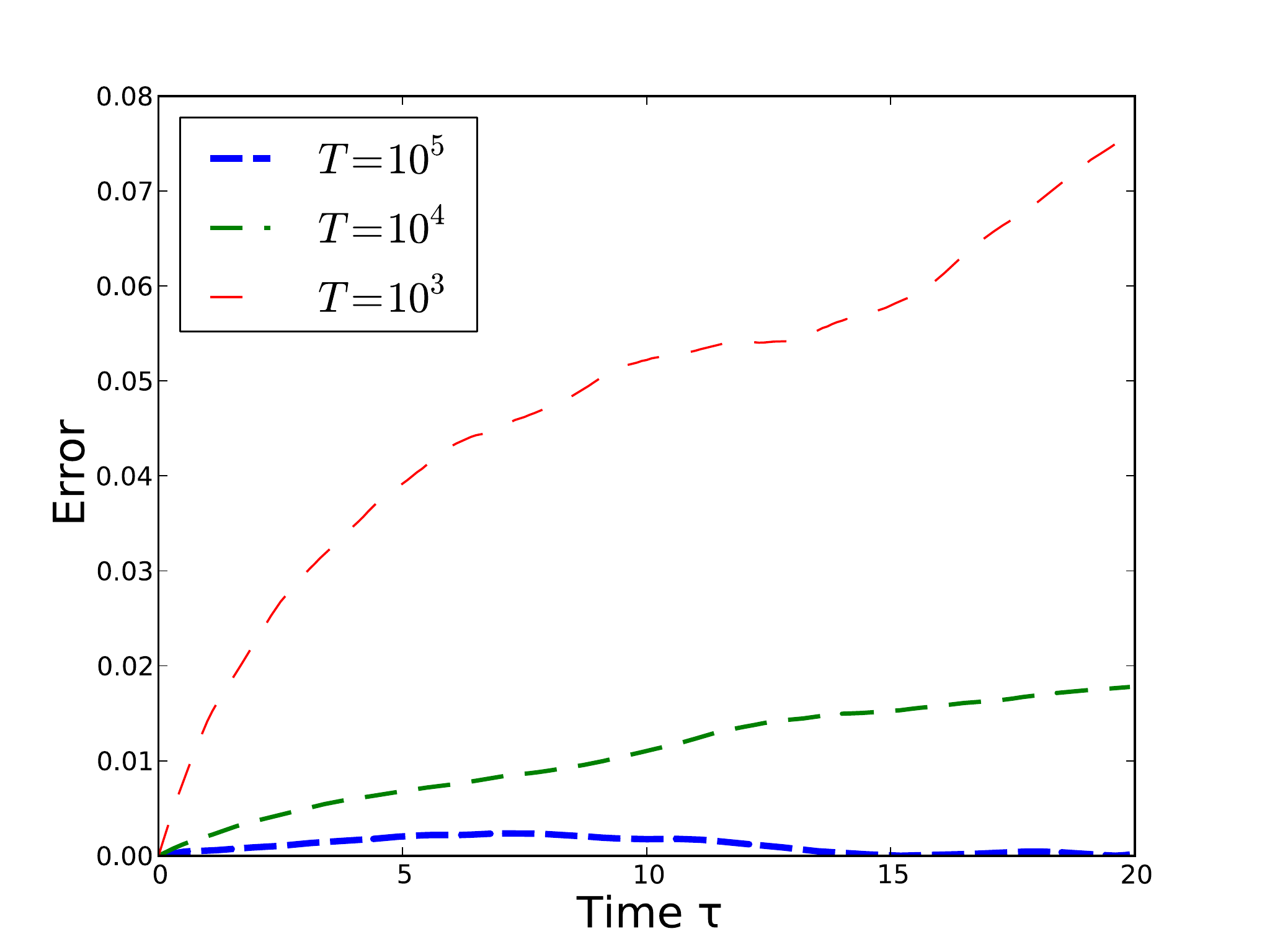} %GAA_diff_log.pdf}
  	\label{fig:GAA_purediff_sum-b} }
\caption{(a) The autocorrelation of the simulated fluorescent 
signal, $G_{AA}(\tau)$, for pure 3D diffusion with twenty five $A$
particles. The motions of all these particles are completely
statistically independent. Here, $\epsilon$ = 0.1, $\omega=10\epsilon_A$
$\Delta t$ = 0.1, and total 
time steps in the simulations are 
$T=10^5$ (blue dash). 
The black solid line is EM's analytical result\cite{elson1974fluorescence} (\ref{app:pdG}); 
the thick dashed line (blue) is the simulation result, and the thin dashed lines (red) are the
simulation error bars calculated from the standard deviation calculated over 30 realizations.
(b) The absolute value of the difference between the 
simulated $G_{AA}(\tau)$ and the EM's results for different total time steps: $T = 10^3$ (red dash), $10^4$ (green dash), and
$10^5$ (blue dash). The maximum error for $T=10^5$ is $2.3\times10^{-3}$ and for $T=10^3$ is $7.6\times10^{-2}$.
}
\label{fig:GAA_purediff}
\end{figure}

\begin{figure}
\centering
  \subfigure[]{
  	\includegraphics[width=0.45\paperwidth]{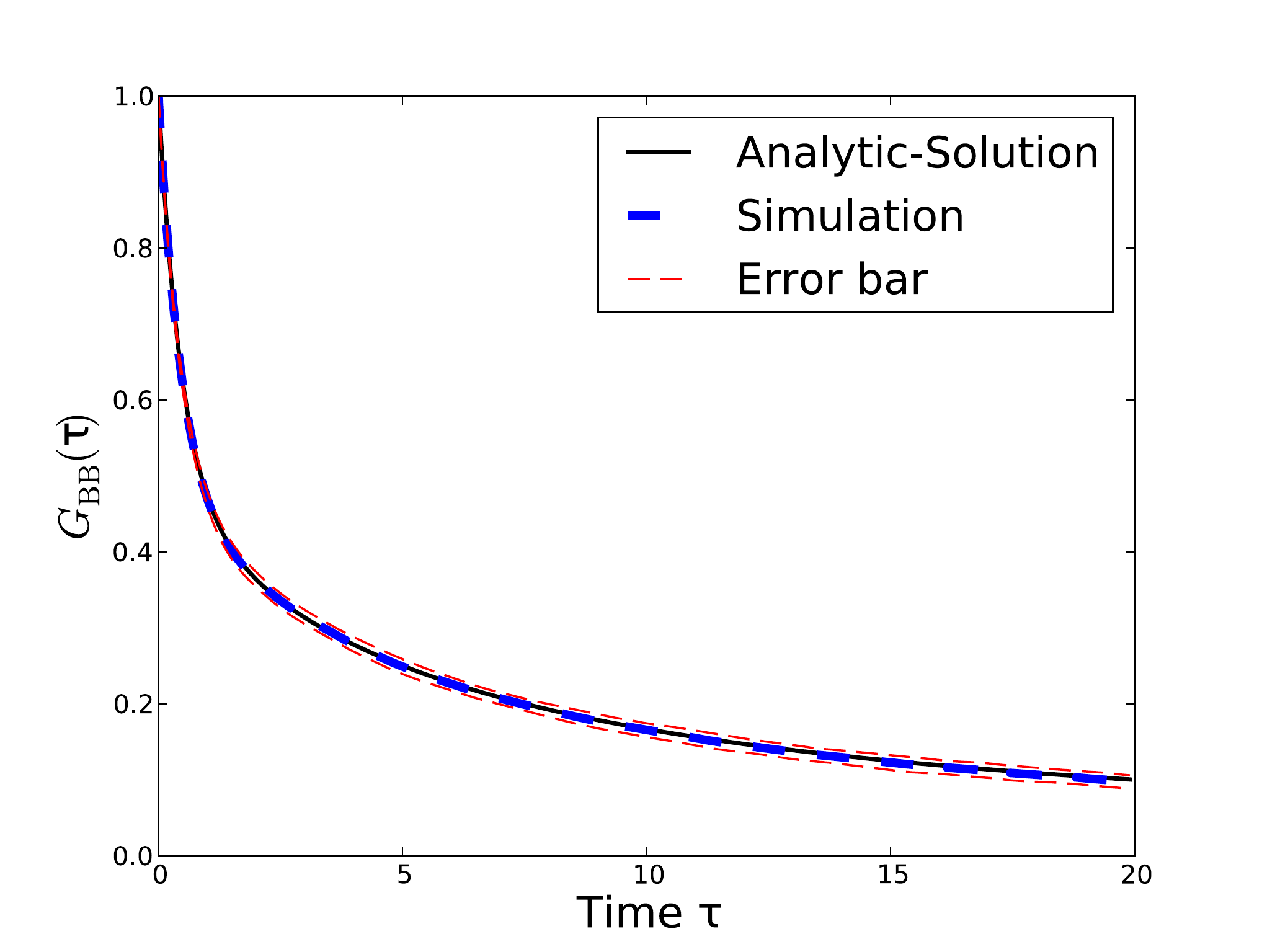} %GBB_sum.pdf}
  	\label{fig:GAA_purediff_sum-a} }
  \subfigure[]{
        \includegraphics[width=0.45\paperwidth]{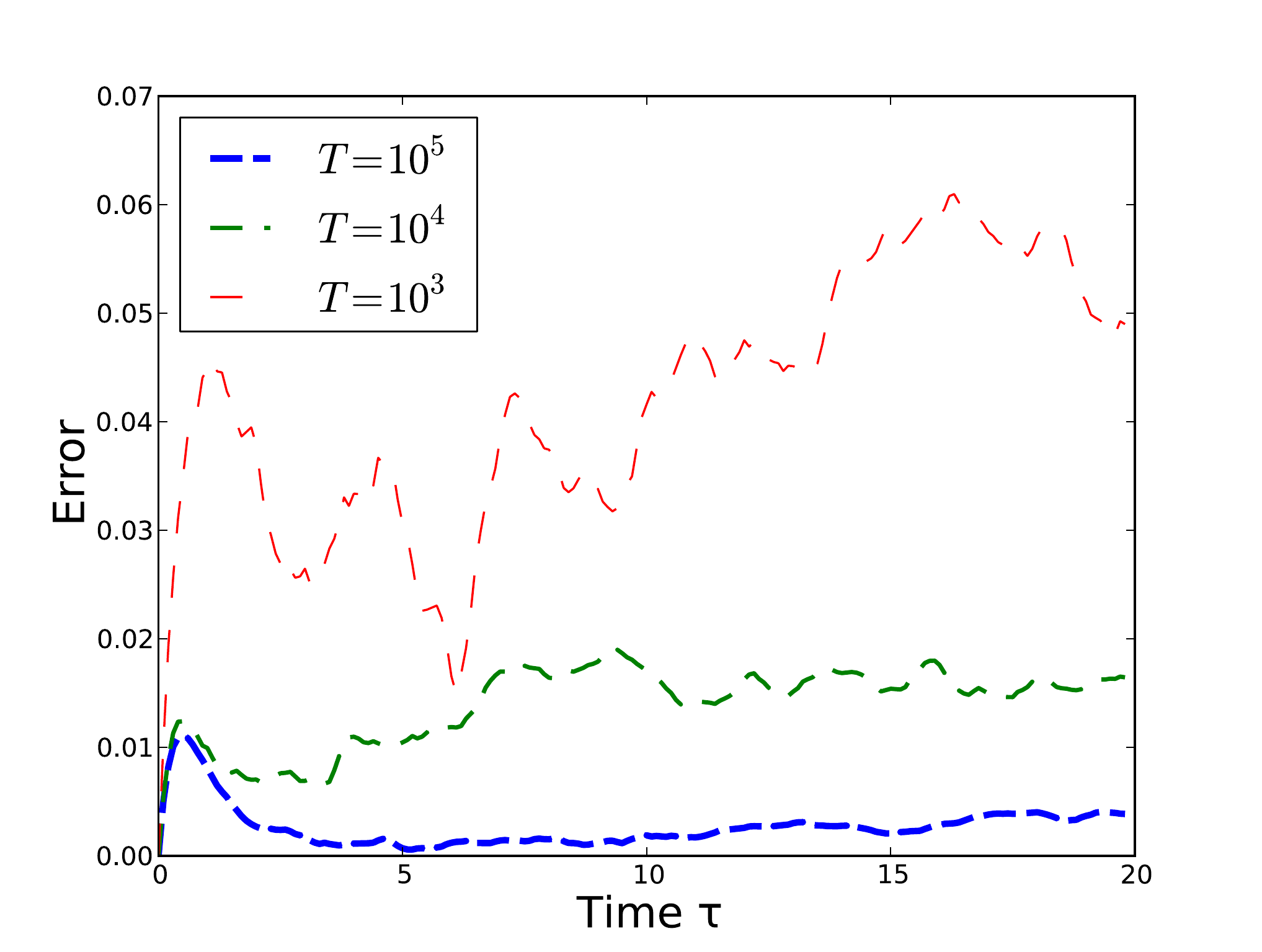} %GBB_diff.pdf}
  	\label{fig:GAA_purediff_sum-b} }
\caption{(a) The temporal autocorrelation of the photocurrent $G_{BB}(\tau)$ for particle $B$ in the unimolecular isomerization, $A\xrightleftharpoons[k_b]{k_f} B$. Here, $\epsilon_A=\epsilon_B=0.1$, $\omega=10\epsilon_A$, $\Delta t=0.1$, $T=10^5$ and $k_f=k_b$. The black solid line is EM's analytical result\cite{elson1974fluorescence} (\ref{app:urwd}); the thick dashed line (blue) is the simulation result, and the thin dashed lines (red) are the error bars from standard deviation calculated over 30 realizations. (b) The difference between the analytical solution and simulation results of the unimolecular isomerization for $G_{BB}(\tau)$ at different total measurement times, $T=10^5$, $T=10^4$ and $T=10^3$. The maximum error for $T=10^5$ is $1.1\times10^{-2}$ and for $T=10^3$ is $6.0\times10^{-2}$.}
\label{fig:GBB_ABonly_sum}
\end{figure}

\noindent
% 	Methods for Monte Carlo simulations are given in 
% Sec. \ref{sec:methods}.  
As controls, both pure diffusion
and unimolecular reaction with diffusion are carried out
first.  The EM theory for both these linear stochastic dynamics
is exact; hence, it is expected to agree completely with 
the simulations, aside from statistical uncertainties.
This is indeed the case as shown in Figs. (\ref{fig:GAA_purediff}a) and (\ref{fig:GBB_ABonly_sum}a). 

For the nonlinear reaction with diffusion, EM theory has no exact solution; however, 
linearization around equilibrium yields a correlation curve in the form of an integral.
This integral can be approximated asymptotically for certain range of parameters, or it can
be solved numerically. For details on the analytic solutions, see the Appendix \ref{sec:Appendix}. For certain values of the equilibrium concentration, the simulation deviates from these solutions. We account this deviation to the non-linear effects that are
present in the simulation but not in EM theory. The maximum errors are calculated with the maximum norm, which for the error between $\mathbf{x}=(x_1, \cdots, x_n)$ and $\mathbf{y}=(y_1,\cdots, y_n)$ is given by $||\mathbf{x}-\mathbf{y}||_{\textrm{max}} = \textrm{max}\{|x_1-y_1|, \cdots, |x_n-y_n|\}$. 

% Note this is done over a discrete set of points $x_i$, which separation is given by the computational resolution of our curves.  
% The asymptotic approximation of the integral yields accurate results only for certain values of the parameters,
% while its numerical approximation yields good results for any parameters. 
To ensure the validity of the results, two asymptotic
limits, where the nonlinear effects are known to be negligible, are employed as a third and fourth control. In order to provide a quantification of the non-linear deviation, its magnitude is studied as a function of the forward reaction rate and the number of molecules. 
All these will be further discussed in detail. \\

\noindent
{\bf\em Control I: pure diffusion.}
The calculated $G_{AA}(\tau)$, normalized autocorrelation
function, from simulations with the number of particles, $N_A=25$, and total measurement time, $T=10^5$, is 
plotted along the analytical solution in Fig. (\ref{fig:GAA_purediff}a). Aside from statistical errors, the simulation results
agree completely with the analytical solution (\ref{app:pdG}) as expected. We further calculate the difference between the analytical solution and the simulation results with different total measurement time, as depicted in Fig. (\ref{fig:GAA_purediff}b). Consistently, it shows that this difference is decreasing with increasing total measurement time $T=N_t\Delta t$. Note that since particles are treated independently, an increase in the total measurement time is equivalent to increasing the number of particles.\\ 

\noindent
{\bf\em Control II: unimolecular reaction with diffusion.}
Analogously to the pure diffusion case, we calculate the temporal autocorrelation of the photocurrent, $G(\tau)$, for either species of particles. Without loss of generality, we choose the initial number of $A$ and $B$ particles equal, i.e. $N_A=N_B$. We calculate the photocurrent for particle $B$ along with the analytic solution (\ref{app:urwd}), as plotted in Fig. (\ref{fig:GBB_ABonly_sum}a). The simulation results are again in good agreement with the analytic solution. In addition, we calculate the difference between the analytical solution and simulation results of $G_{BB}(\tau)$ for different total measurement times. As depicted in Fig. (\ref{fig:GBB_ABonly_sum}b), the error between the analytical result and the simulation decreases as the total measurement time increases. Once more, as the particles react independently of each other, an increase in the total measurement time is equivalent to increasing the number of particles evenly.\\

% (MAX ERROR $4.3\times10^{-3}$)
%

%

%
% \begin{figure}
% \centering
% %\includegraphics[scale=0.3]{Figures/GBB_diff.eps}
% %\includegraphics[scale=0.3]{Figures/GBB_diff_log.eps}
%   	\includegraphics[width=0.45\paperwidth]{Figures/Fig_4_UnimErrors.pdf} %GBB_diff.pdf}
% \caption{The difference between the analytical solution and simulation results of the unimolecular isomerization for $G_{BB}(\tau)$ at different total measurement times, $T=10^5$, $T=10^4$ and $T=10^3$. The maximum error for $T=10^5$ is $1.1\times10^{-2}$ and for $T=10^3$ is $6.0\times10^{-2}$. }
% \label{fig:GBB_ABonly_diff}
% \end{figure}
%

\begin{figure}
\centering
\includegraphics[width=0.45\paperwidth]{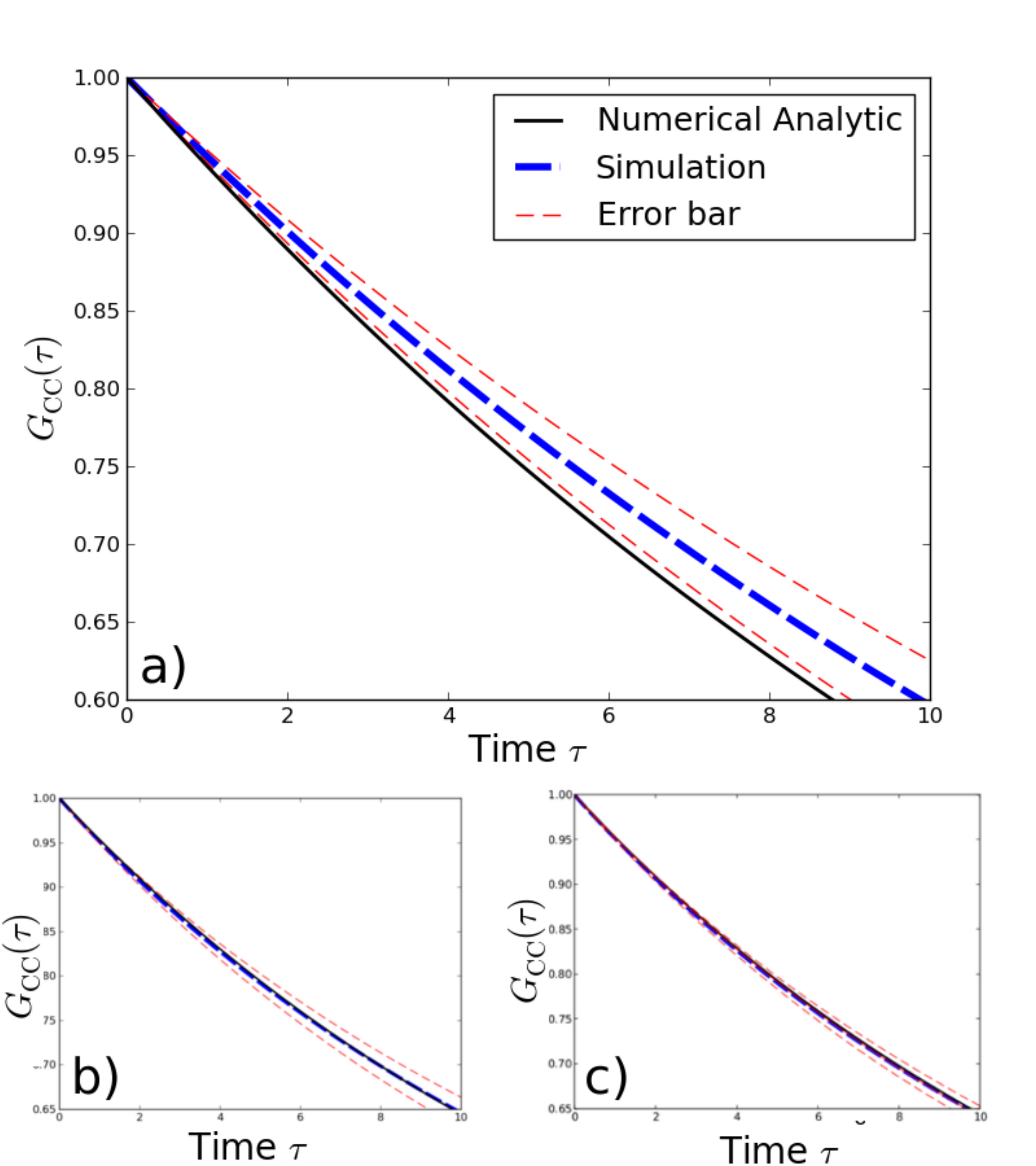}
%\includegraphics[scale=0.3]{Figures/GCC_sum_log.eps}
%   \subfigure[]{
%   	\includegraphics[width=0.45\paperwidth]{Figures/Fig_5_BimDiffComp.pdf} %GCC_sum.pdf}
%   	\label{fig:GAA_purediff_sum-a} }
%   \subfigure[]{
%   	\includegraphics[width=0.45\paperwidth]{Figures/Fig_5_Zoom_BimDiff.pdf} %GCC_sum_log.pdf}
%   	\label{fig:GAA_purediff_sum-b} }
\caption{a) The temporal autocorrelation of the photocurrent, $G_{CC}(\tau)$, for particle $C$ in the bimolecular reaction with $N_A$, $N_B$ and $N_C$ small and of the same order (around $35$ each). The thick dashed line (blue) is the simulation result, and the thin dashed lines (red) are the error bars given by the standard deviation calculated over 30 realizations. Here $\epsilon_A=\epsilon_C= 0.0033$, $\epsilon_B=0.0066$, $\omega=25\epsilon_B$, $\Delta t=0.1$, $T=2\times10^6$ and $R=0.05$. The calculated unbinding radius for this case was in average $R_u = 0.302$. The maximum error between the simulation and the analytic solution is around $0.04$. b) Same parameters as in a) but with the asymptotic limit of many ligands $N_B$. c) Same parameters as in a) for the asymptotic limit of large number of molecules $N_A,N_B,N_C\gg 1$. The full plots of the asymptotic limits plots b) and c) are shown in Figs \ref{fig:GCC_ABC_lim} and \ref{fig:GCC_ALL_lim}.  } %, and $R_A=R_B=0.05$.}
\label{fig:GCC_ABC_sum}
\end{figure}

\noindent
{\bf\em Nonlinear reaction with diffusion.}
We choose $N_A,N_B$ and $N_C$ small and of the same order. We calculate the temporal autocorrelation of the photocurrent $G(\tau)$ for particle $C$ and compare it against the analytical solution. In Fig. (\ref{fig:GCC_ABC_sum}a)  two solutions for the correlation curve are shown: the simulation curve and the numerical approximation of the analytic solution integral (\ref{app:nrwd}). As $N_A,N_B$ and $N_C$ are of the same order, the asymptotic approximation of the integral is outside its range of validity (see Appendix \ref{sec:Appendix}), and it's not included in the plot. A noticeable disagreement between the simulation and the numerical analytic integration result is observed as depicted in Fig. (\ref{fig:GCC_ABC_sum}a). The maximum error between the simulation and the analytic solution is around $0.04$, and the theory will produce an error of $11.5 \%$ in the forward reaction rate if fitted to the central tendency in the simulation curve. The plots in Figs. (\ref{fig:GCC_ABC_sum}b) and (\ref{fig:GCC_ABC_sum}c) correspond to the two asymptotic limits where the nonlinear effects are negligible and the simulation converges to the analytic linear theory. These two limits will be discussed next. \\
%_A=\Delta t_B=\Delta t_C=

\begin{figure}
\centering
  \subfigure[]{
	\includegraphics[width=0.45\paperwidth]{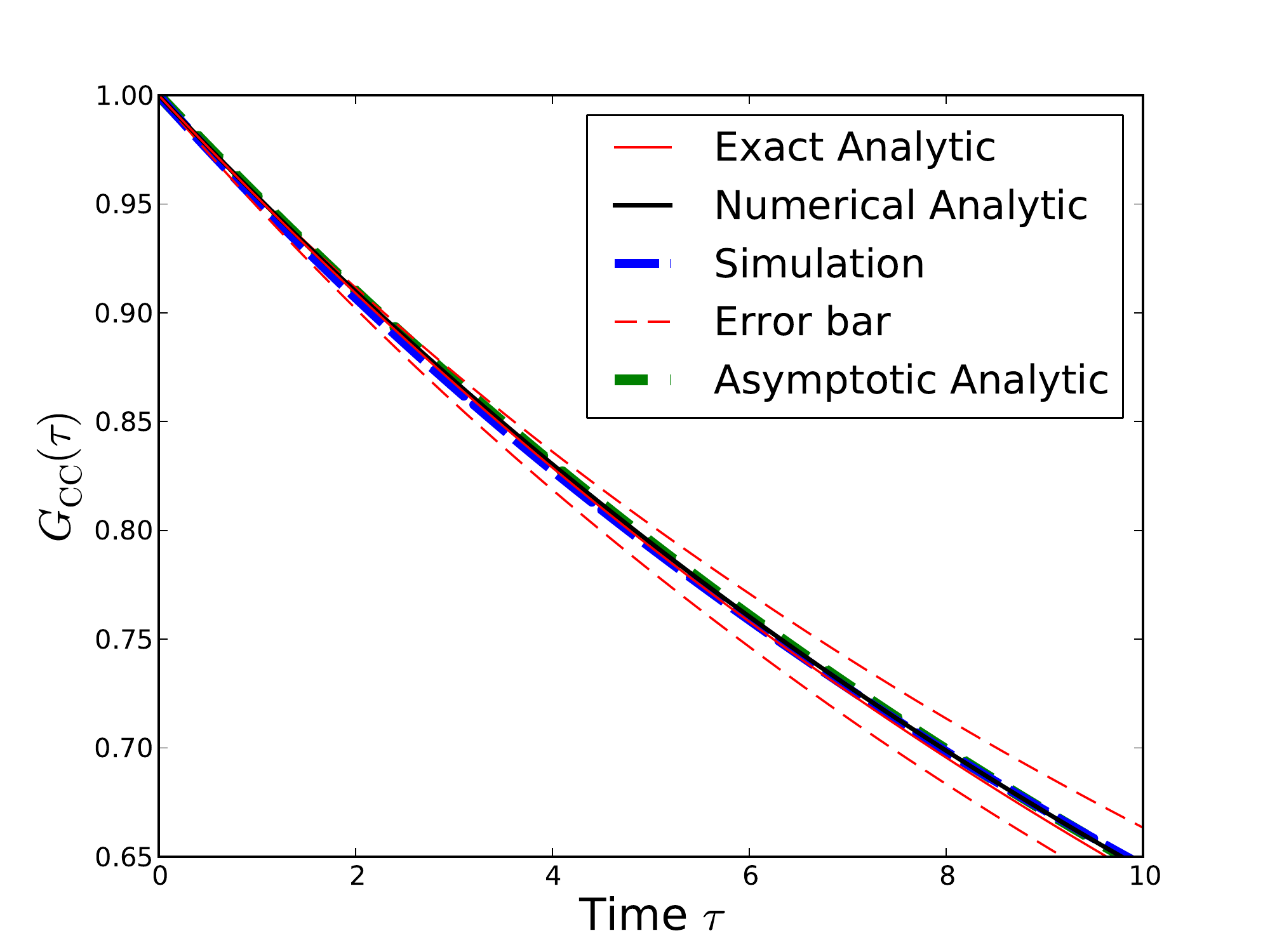}
  	\label{fig:GAA_purediff_sum-a} }
  \subfigure[]{
        \includegraphics[width=0.45\paperwidth]{BimLimZoom.pdf}
  	\label{fig:GAA_purediff_sum-b} }
\caption{a) The temporal autocorrelation of the photocurrent, $G_{CC}(\tau)$, for particle $C$ in the bimolecular reaction with $N_B\gg N_A,N_C$; $N_B \thicksim 2000$, $N_A,N_C \thicksim 20$. The error bars are once again given by the standard deviation calculated over 30 realizations. Here $\epsilon_A=\epsilon_C= 0.0033$, $\epsilon_B=0.0066$, $\omega=25\epsilon_B$, $\Delta t=0.1$, $T=2\times 10^6$ and $R=0.05$. The calculated unbinding radius for this case was in average $R_u = 0.414$. The maximum error between the numerical analytic solution and the simulation is $3.2\times10^{-3}$. b) Zoomed in version of a).}
\label{fig:GCC_ABC_lim}
\end{figure}

\noindent
{\bf\em Asymptotic limit I: large number of ligands}.
In the case where $N_B\gg N_A,N_C$, the concentration of $B$ molecules barely fluctuates. Consequently, the nonlinear reaction with diffusion approaches asymptotically a linear unimolecular reaction with diffusion, i.e. the nonlinear effects are negligible. A simple example is given by the law of mass action for $A+B\xrightleftharpoons[k_b]{k_f} C$. 
If $b$ barely fluctuates around $b_0$, the concentration of $C$ follows
\begin{align*}
 \frac{dc}{dt} = K_f a - k_b c, \hspace{5mm} \mathrm{with:} \hspace{5mm} K_f = k_f b_0,
\end{align*}
with $a,b,c$ the concentrations of $A,B,C$ and $k_f$ the second order rate constant.
The last equation is clearly linear and EM theory provides an exact result for it. In Fig. \ref{fig:GCC_ABC_lim}, besides the same two solutions as plotted before, the exact analytic solution approached as $N_B$ becomes $N_B\gg N_A,N_C$ is also included. Since it's in its range of validity and the non-linear effects are negligible, the asymptotic approximation of the linear analytic solution (\ref{app:nrwd}) is also plotted. The four solutions, including the simulation and the numerical integration of (\ref{app:nrwd}), converge to the same correlation curve, as expected. This result is clearly depicted in Fig. \ref{fig:GCC_ABC_lim}. It also shows consistency  with our two controls, since the error decreases when increasing the number of particles. Recovering the correct asymptotic limit when $N_B\gg N_A,N_C$ serves as a third control to validate our simulation. \\

\begin{figure}
\centering
  \subfigure[]{
  	\includegraphics[width=0.45\paperwidth]{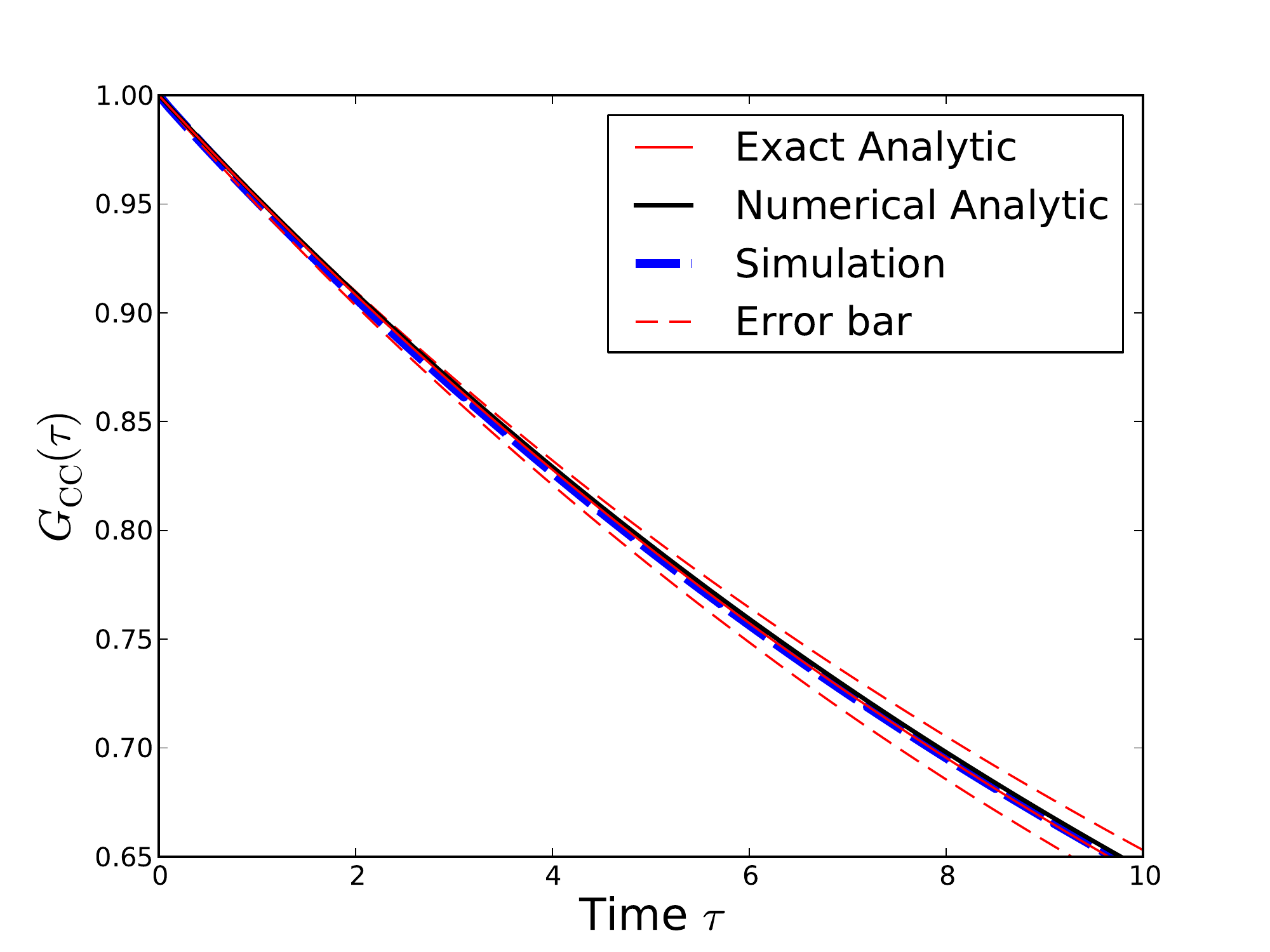} %GCC_sum.pdf}
  	\label{fig:GAA_purediff_sum-a} }
  \subfigure[]{
  	\includegraphics[width=0.45\paperwidth]{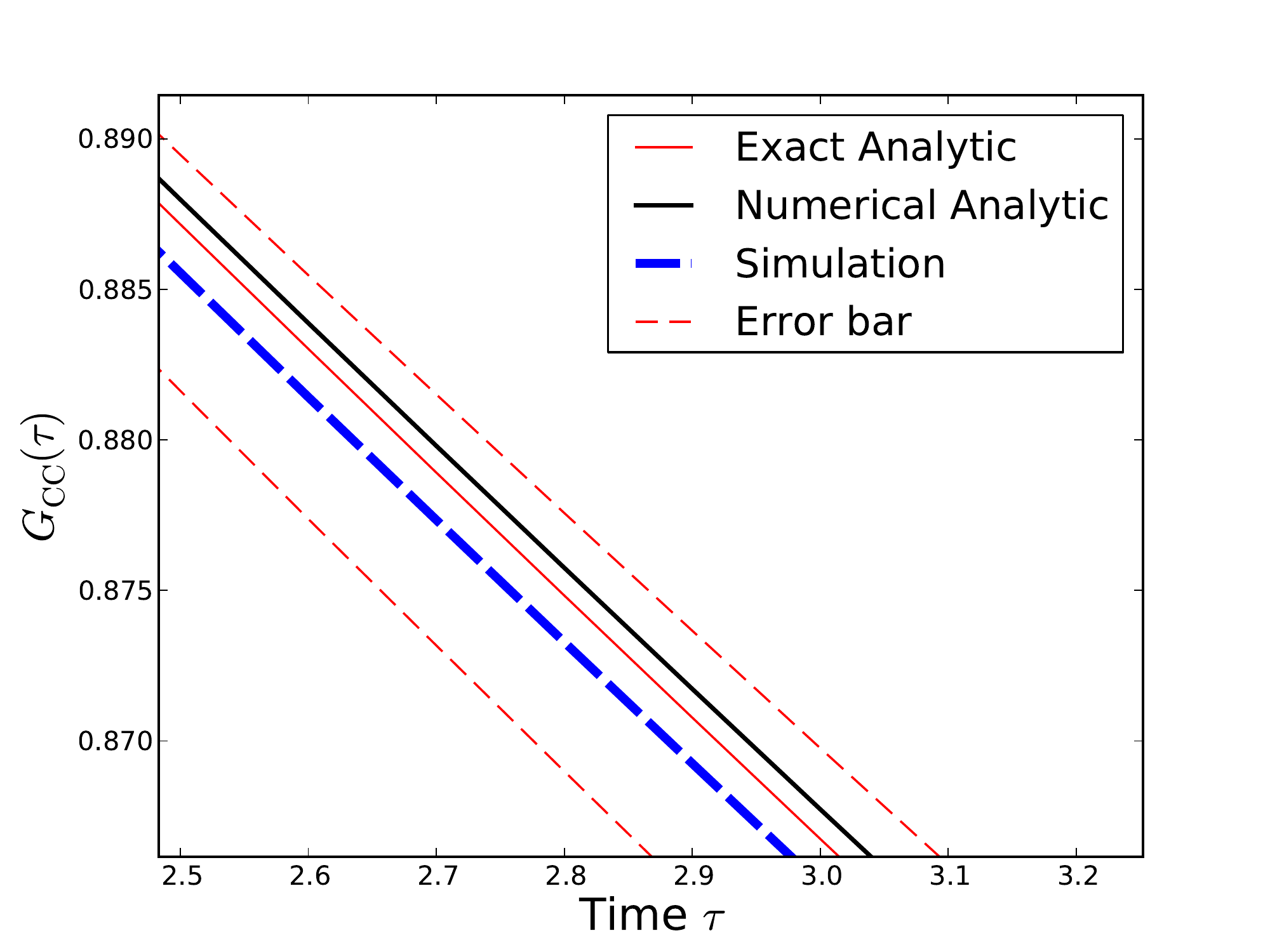} %GCC_sum_log.pdf}
  	\label{fig:GAA_purediff_sum-b} }
\caption{a) The temporal autocorrelation of the photocurrent, $G_{CC}(\tau)$, for particle $C$ in the bimolecular reaction with $N_B,N_A,N_C\gg 1$; $N_A,N_B,N_C \thicksim 500$. The error bars are once again given by the standard deviation calculated over 30 realizations. Once again the parameters are $\epsilon_A=\epsilon_C= 0.0033$, $\epsilon_B=0.0066$, $\omega=25\epsilon_B$, $\Delta t=0.1$, $T=2\times 10^6$ and $R=0.05$. The calculated unbinding radius for this case was in average $R_u = 0.132$. The maximum error between the numerical analytic solution and the simulation is $2.8\times10^{-3}$. b) Zoomed in version of a).}
\label{fig:GCC_ALL_lim}
\end{figure}

\noindent
{\bf\em Asymptotic limit II: large number of all molecules}.
If the number of all the molecules is increased, i.e. $N_A,N_B,N_C\gg 1$, the concentration of any of the species barely fluctuates around equilibrium. Analogously to
the previous case, the nonlinear effects become second order and the system approaches a linear solution, which is plotted as the exact analytic solution in Fig. \ref{fig:GCC_ALL_lim}. Additionally, the numerical analytic solution and the simulation curves are plotted as before. In this case, the asymptotic approximation of the linear analytic solution (\ref{app:nrwd}) is not in its range of validity, so it is not included. As we can see in Fig. \ref{fig:GCC_ALL_lim}, the simulation, the numerical analytic solution  and the exact linear unimolecular solution are all converging, i.e. the non-linear effects are becoming negligible as the number $N_A,N_B$ and $N_C$ is increased. This asymptotic limit serves as a fourth control to validate the simulation.\\

\begin{figure}
\centering
  \subfigure[]{
  	\includegraphics[width=0.43\columnwidth]{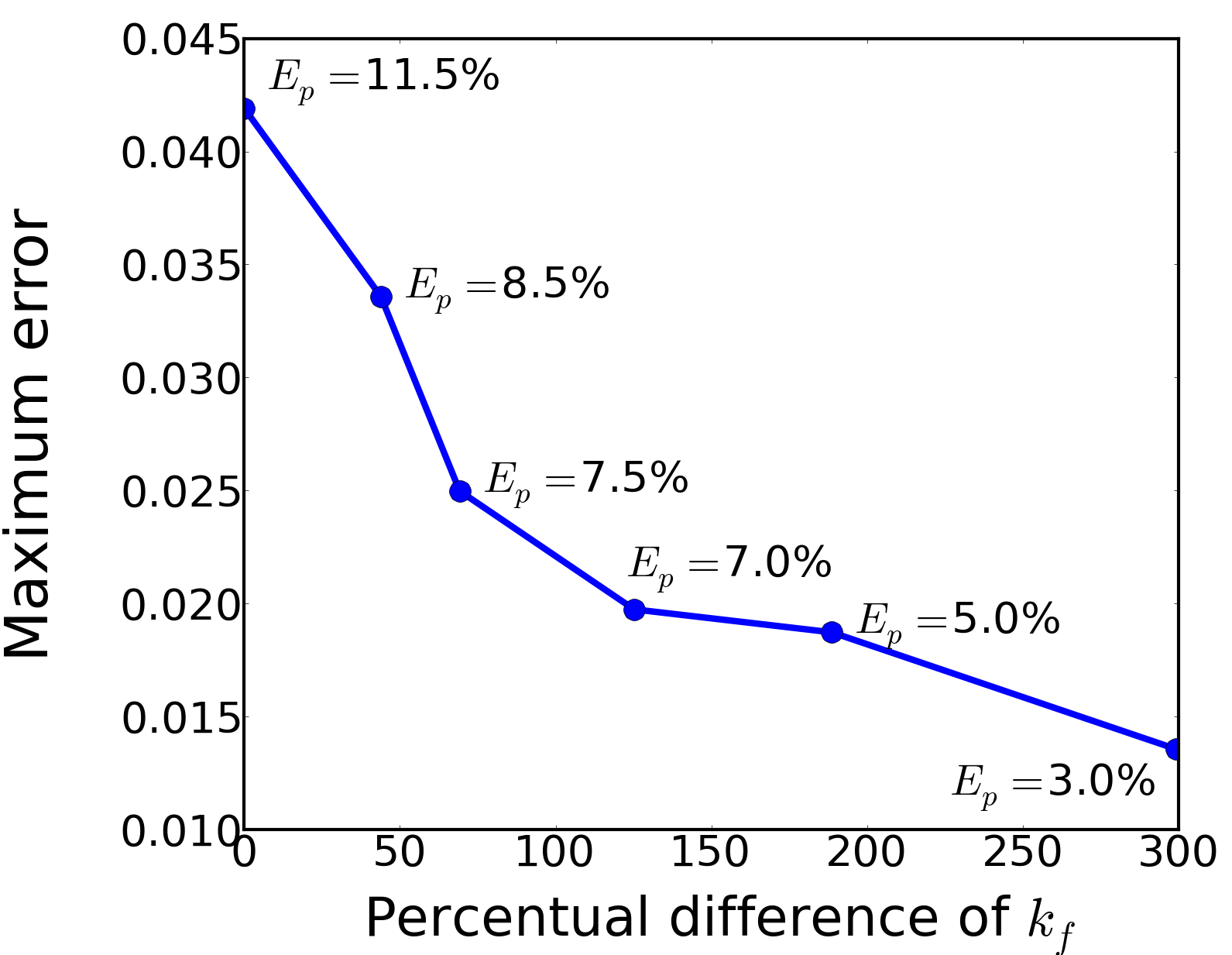} %GCC_sum.pdf}
  	\label{UPa} }
  \subfigure[]{
  	\includegraphics[width=0.43\columnwidth]{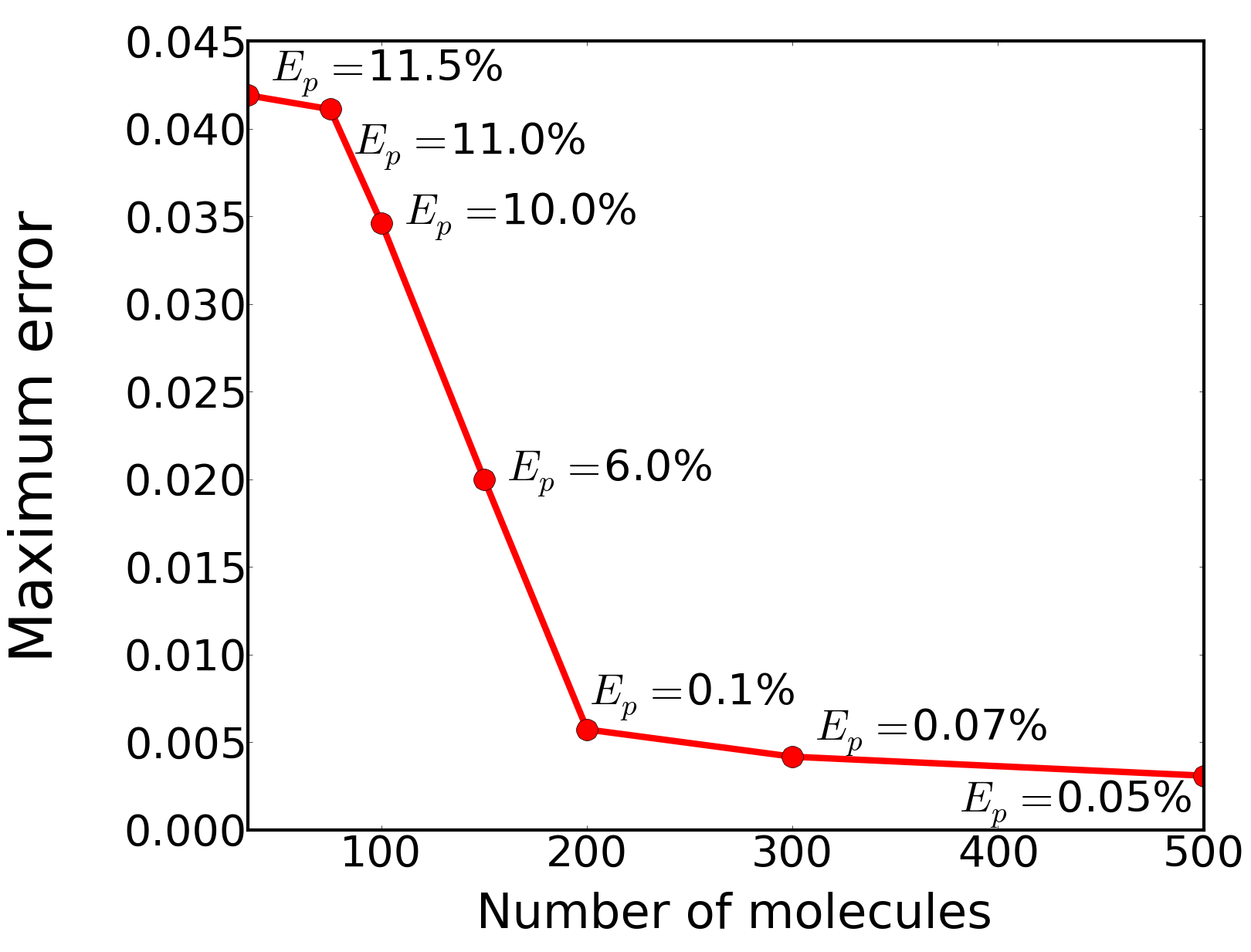} %GCC_sum_log.pdf}
  	\label{UPb} }
\caption{Plot of the maximum error between the numerical analytical correlation curve and the simulated one as a function of: a) Percentage deviation between forward reaction rates, b) Number of molecules $N=N_A,N_B,N_C$. The first point on the left of both plots corresponds to the simulation with $N_A,N_B$ and $N_C\thicksim 35$ plotted in Fig. \ref{fig:GCC_ABC_sum}. The percentage deviation in a) is in reference to this point. The last point to the right in b) corresponds to the case plotted in Fig. \ref{fig:GCC_ALL_lim}. Each point in the curves has an $E_p$, corresponding to the percentage error when predicting the forward rate $k_f$ by fitting the theory to the simulation central tendency.}
\label{fig:GCCdepend}
\end{figure}

\noindent
{\bf\em Nonlinear deviation variation}.
In order to better quantify the deviation for the nonlinear reaction case shown in Fig. \ref{fig:GCC_ABC_sum}. We calculated how this deviation varies in term of the relative change in the forward reaction rate (Fig. \ref{UPa}) and with the uniform increase in the number of all molecules (Fig. \ref{UPb}).

The plot shown in Fig. \ref{UPb} shows the convergence to the asymptotic limit II: Large number of all molecules. As expected from our previous result, the error is gradually reduced as the number of all molecules is increased. Furthermore, on the plot in Fig. \ref{UPa}, we manipulated the forward reaction rates $k_f$. As these rates are related to the diffusion coefficients $D_\xi = \epsilon_\xi/6\Delta t$ by equation (\ref{frate}), the best way to manipulate the reaction rate is to modify the characteristic length step $\epsilon_\xi$. In this case we multiplied $\epsilon_\xi$ (with $\xi=A,B,C$) by a constant between one and two, which yields a percentage variation of the forward reaction rate up to $300 \%$. 

The deviation in Fig. \ref{UPa} is reduced as the reaction rate is increased. This is expected from the law of mass action for $A+B\xrightleftharpoons[k_b]{k_f} C$, where their concentrations $a,b$ and $c$ satisfy $(ab) k_f/k_b = c$ in the steady state. In this expression it is clear how an increase in the forward rate $k_f$ can be equivalent to an increase of $b$. Consequently this case is equivalent to the asymptotic limit I: large number of ligands. As the characteristic length step is increased, the error by approximating diffusion by a random walk is also increased. However, this error is still not relevant as the overall error shown in Fig. \ref{UPa} decays as the reaction rate is increased.     

Also note each point in Fig. \ref{fig:GCCdepend} has an associated percentual error, $E_p$. This was obtained by testing slightly different values for the forward rate $k_f$ in the analytic curve until the maximum error against the simulation was below $0.002$. Ideally a least squares fitting could be employed. However, we are currently developing a faster and more robust simulation to illustrate a complete landscape of the deviations as a function of two or more parameters, which will include a least square fitting to obtain the percentual error. The results will be provided in an upcoming manuscript.

\section{Discussion}
\label{sec:discuss}
\noindent

% In this work, we use a $10\times10$ box as the simulation domain and the periodic boundary condition is employed. We find that 50 is an appropriate value for the ratio of box size to the length step ($\epsilon$). If a smaller value is adopted, the periodic box will not be a good approximation for an infinite domain any more. On the other hand, if a larger value is used, the particles distribute in the box too sparsely and the value of Gaussian intensity (see eq. ~\ref{equ:laser_intensity}) decreases exponentially. When the $G(\tau)$ is normalized by $G(0)$, an infinitely small denominator is found.
% 
% In the calculation of each normalized $G(\tau)$, 200 runs were performed to get the ensemble average. All the units in the present work are reduced model units. 

We showed that our simulation recovers the correct reaction rates of the Smoluchowski's model for reversible reactions providing consistency between the simulation parameters and the parameters used by EM theory. As first controls, the exact solutions of EM theory for the pure diffusion and the unimolecular reaction with diffusion were recovered by our simulation. In addition, to further validate the simulation, two asymptotic limits were tested. In the asymptotic limits of large number of ligands and large number of molecules, the simulation and the theoretical correlation curves converge as expected. This confirms that in these limits the nonlinear effects become negligible and linear EM theory is very accurate. However, in the case where
$N_A,N_B$ and $N_C$ are not very big and have similar values (around 35 each), we showed the simulation results deviate from those of linear EM theory. As the simulation proved to be an accurate model, we account the deviations due to nonlinear effects that are depreciated in linear EM theory\cite{elson1975concentration}, as briefly shown in the Appendix \ref{sec:Appendix}. 

The dependance of the deviation in the number of molecules and the reaction rates was analyzed. When employing an increasing number of all of the molecules, we observe the deviation is reduced and the system gradually reaches the asymptotic limit II, as expected. When the forward reaction rate was gradually increased up to $300\%$ its original value, we again observe the error is reduced. This is due the fact that increasing the reaction rate produces a similar effect to that of the increase in the number of ligands, which corresponds to the asymptotic limit I, where the error is also reduced.  

For most experimental scenarios the assumptions $N_B\gg N_A,N_C$ or $N_A,N_B,N_C\gg 1$ are appropriate, and the results provided by EM theory are very accurate. Nonetheless, the present paper showed that when the number of molecules is small, nonlinear effects are not negligible showing a deviation between the simulation of the nonlinear model and linear EM theory. However, experimental FCS correlation curves usually involve other sources of noise not considered in the simulated fluctuations, like mismatch of refractive indexes and photobleaching of fluorophores amongst others. As we found the deviation to be small with a $11.5\%$ error in the reaction rates, it is likely that the error produced by the nonlinearity is still within the experimental uncertainty of current laboratory measurements, so it might require a more carefully designed setup and clean system to test our theory. 

In particular, we do not believe the nonlinear effect will be relevant in the current cellular biophysical investigations. Rather, the significance of the present work is to bring quantitative experimental measurements on nanometric, nonlinear chemical reactions a step closer to a stochastic theoretical framework. It also intends to call attention to nonlinear kinetics in the fluctuation chemistry setting. Certainly, we hope this work paves the way to study more complex nonlinear reactions with concentration fluctuations. We are certain for other more complex reaction systems the nonlinear effect can be larger. Furthermore, on the practical side, nonlinear chemical reactions in biology are widely present; so being able to show, at least in the simplest case, the EM theory works well, even in a context where nonlinearities are significant, is a relevant contribution for currenct practice of FCS in biochemistry.

With the increasing accuracy and ability of single-molecule techniques, results like the ones obtained herein will become experimentally accessible. We also believe FCS will grow more and more into an analytical tool. In a clear analytical chemistry setting, an $11\%$ deviation would be significant . It is true that such level of quantification is still in development; but we hope results like ours provide additional motivation. Developing nonlinear chemical reaction theory with fluctuations, in terms of FCS or more generally speaking, remains a challenge.

% Although the deviations are small and most likely not relevant experimentally due to toher sources of noise

%Acknowledgements: Elliot Elson, Wei Min, Saveez Saffarian, Michelle Wang, Claus Seidel, Jie Liang, A. Szabo

\section{Appendix: EM theory analytic results}
\label{sec:Appendix}
\noindent
Three main results from EM theory \cite{elson1975concentration} are incorporated in this
appendix. For the three cases, pure diffusion, unimolecular isomerization and nonlinear
reaction with diffusion, the full expressions for the correlation curves are given. Additional details 
are given to explain why EM theory is a linear theory for the bimolecular reaction. 

The parameters used are $\Delta t$ the time step, $\epsilon_\xi$ with $\xi=A,B$ or $C$
the diffusion length step for every time step taken, $D_\xi$ the diffusion coefficient of $\xi$
and $\omega$ focal volume given by the radius of the Gaussian laser. \\

\noindent
{\bf\em Pure diffusion.}
The normalized auto correlation curve for purely diffusive $A$ molecules is given by, 
\begin{gather}
 G_{AA}(\tau) = \frac{1}{1+\tau/\tau_{D_A}} \label{app:pdG}\\
 \mathrm{with} \hspace{5mm} \tau_{D_A} = \frac{\omega^2}{4 D_A} \hspace{5mm} \mathrm{and} \hspace{5mm} D_A = \frac{\epsilon_A^2}{6 \Delta t}.
 \label{app:pd}
\end{gather}

\noindent
{\bf\em Unimolecular reaction with diffusion.}
The normalized autocorrelation curve of $B$ for the diffusive reaction $A\xrightleftharpoons[k_b]{k_f}B$ with $D_A=D_B$ is given by,
\begin{gather}
 G_{BB}(\tau) = \frac{1}{1+K_{eq}}\left(\frac{K + \exp{(-R\tau)}}{1+\tau/\tau_{D_B}}\right),
 \label{app:urwd}
\end{gather}
with $R=k_f + k_b$, $K_{eq} = k_f/k_b$ and,
\begin{gather*}
 \tau_{D_B} = \frac{\omega^2}{4 D_B} \hspace{5mm} \mathrm{and} \hspace{5mm} D_B = \frac{\epsilon_B^2}{6 \Delta t}.
\end{gather*}

\noindent
{\bf\em Nonlinear reaction with diffusion.}
The normalized autocorrelation curve of $C$ for the diffusive reaction $A+B\xrightleftharpoons[k_b]{k_f}C$ with
$D_A=D_C=D$ is given by the integral,
\begin{equation}
  G_{CC}(\tau) = \frac{\omega^2}{4 \pi} \int_{-\infty}^\infty \int_{-\infty}^\infty \mathcal{F}(\tau,\nu_x,\nu_y) d\nu_x d\nu_y,
 \label{app:nrwd}
\end{equation}
with
\begin{equation*}
 \mathcal{F}(\tau,\nu_x,\nu_y)=\exp{\left[-(\nu_x^2+\nu_y^2)\frac{\omega^2}{4}\right]} Z_{CC}(\tau,\nu_x,\nu_y),
\end{equation*}
where $\nu_x$ and $\nu_y$ are the Fourier frequency variables corresponding to the spatial variables $x$ and $y$.
The integral cannot be solved exactly; nonetheless, we have two possible approaches: provide an asymptotic approximation or solve it numerically. The former, as done in EM theory\cite{elson1975concentration}, requires that $D_B\gg D$
and $C_A^{eq}\thicksim C_C^{eq}\ll C_B^{eq}\thicksim K_{eq}^{-1}$. The numerical solution doesn't require any of these conditions. In section (\ref{secc:results}), we employ both  approaches to compare the theory against our simulation. 

In equation (\ref{app:nrwd}), $Z_{CC}(\tau,\nu_x,\nu_y)$ is given by 
\begin{gather}
 Z_{jl}(\tau,\nu_x,\nu_y) = \sum_s X_l^{(s)} Y_j^{(s)} \exp{(\lambda^{(s)}\tau)},
\end{gather}
with $j,l=A,B$ or $C$. The quantities $X^{(s)}$ for $s=1,2,3$ are the right eigenvectors, $Y^{(s)}$ are the left eigenvectors and $\lambda^{(s)}$ are the three eigenvalues of matrix $\mathbf{M_0}$ from (\ref{FCSeq2}).
The matrix $\mathbf{M_0}$ is obtained from the reaction diffusion equation for the concentration of the three molecules at position $\textbf{r}$ and time $\tau$. The concentration for the three molecules will be given by the vector $\textbf{C}(\textbf{r},\tau)$ with components ${C_j(\textbf{r},\tau)}$ and $j=A,B$ or $C$. The full \textbf{nonlinear} reaction diffusion equation is given by
\begin{align}
 \frac{\partial \textbf{C}(\textbf{r},\tau)}{\partial \tau} = \textbf{D}\cdot\nabla^2 \textbf{C}(\textbf{r},\tau) + \textbf{M}(\textbf{C}(\textbf{r},\tau)) \cdot\textbf{C}(\textbf{r},\tau),
 \label{FCSeq1}
\end{align}
where $\textbf{D}$ is a diagonal matrix with the $j^{th}$ chemical diffusion coefficients ${D_j}$, and $\textbf{M}$ is the stoichiometry
matrix\footnote{Matrix of reaction coefficients based on the Law of Mass action.}. Note that $\textbf{M}$ is not a constant coefficient matrix, since this is a nonlinear reaction, it depends on the concentration vector $\textbf{C}(\textbf{r},\tau)$. Supposing the system is stationary, the chemical concentrations $\textbf{C}(\textbf{r},\tau)$ 
will reach a thermodynamic equilibrium; therefore, the mean concentration of each component will be given by the ensemble average of the concentration, i.e. 
$\textbf{C}^{eq} = \langle \textbf{C}(\textbf{r},\tau) \rangle $. The ensemble average can be understood as the averaged quantity over many identical systems at a certain time. Although the partial differential equation (PDE) (\ref{FCSeq1}) is deterministic, another \textbf{linear} PDE can be derived for the fluctuations of the system around equilibrium. This fluctuations will be given by $\boldsymbol{\delta C}(\textbf{r},\tau) = \mathbf{C}(\textbf{r},\tau) - \textbf{C}^{eq}$. Substituting into (\ref{FCSeq1}) and dropping the \textbf{nonlinear} terms we obtain the PDE for the fluctuations
\begin{align}
 \frac{\partial  \mathbf{\delta C}(\mathbf{r},\tau)}{\partial \tau} = \textbf{D}\cdot\nabla^2 \mathbf{\delta C}(\textbf{r},\tau) + \mathbf{M_0}\cdot\mathbf{\delta C}(\textbf{r},\tau),
 \label{FCSeq2}
\end{align}
where the constant coefficient matrix $\mathbf{M_0}=$
\begin{align*}
 \left[ \begin{matrix}
 -(\nu^2 D + k_f C_B^{eq}) & -k_f C_A^{eq} & kb \\ 
 -k_f C_B^{eq} & -(\nu^2 D_b +k_f C_A^{eq}) & kb \\
 k_f C_B^{eq} & k_f C_A^{eq} & -(\nu^2 D +k_b) \end{matrix} \right],
\end{align*}
with $\nu^2=\nu_x^2+\nu_y^2$.

The expression for the auto correlation curve (\ref{app:nrwd}) is not trivially obtained from the
calculations just shown, for a full treatment consult EM theory\cite{elson1975concentration}. What is to be noted is that expression (\ref{app:nrwd})
is based on (\ref{FCSeq2}), which is the linearized version of the PDE (\ref{FCSeq1}) around equilibrium.
% As shown in section \ref{secc:results}, the dropped nonlinear terms in (\ref{FCSeq2}) might have 
% relevant contributions for some cases. 

\section*{Acknowledgements}
We thank helpful discussions with Drs. Noam Agmon, Steve Andrews, 
Attila Szabo, and Nancy Thompson. The authors would also like
to thank two anonymous reviewers who greatly improved the quality 
of this paper. MJdR acknowledges partial support
from National Science and Technology Council of Mexico (CONACyT). 
HQ acknowledges partial support from Pacific Northwest National 
Laboratory (PNNL) subcontract  PR 203607.  W.P. and G.L. acknowledge the funding support by the Applied Mathematics Program within the U.S. Department of Energy Office of Advanced Scientific Computing Research as part of the Collaboratory on Mathematics for Mesoscopic Modeling of Materials (CM4), under award number DE-SC0009247. PNNL is operated by Battelle for the
DOE under Contract DE-AC05-76RL01830.

\vskip 0.3cm
% \printthebibliography
% \bibliographystyle{acs}
\bibliographystyle{acm}
\bibliography{refs.bib}

\end{document}